\documentclass[a4paper,12pt]{article}
\usepackage{epsfig}
\usepackage{amsmath}

\newcommand\email[1]{{\it #1}}
\newcommand\institution[1]{{\it #1}}
\newcommand\address[1]{{\it #1}}

\begin{document}
\hyphenation{brems-strah-lung}
\title{Bremsstrahlung in weak charged current polarized
lepton-nucleon deep inelastic scattering}
\author{S. N. Sevbitov and T. V. Shishkina}
\date{\empty}
\maketitle
\vspace{-5mm}
\begin{center}
\institution{Department of Theoretical Physics, Belarusian State University\\}
\address{Nezavisimosti av. 4, Minsk 220030, Belarus\\}
\email{e-mail: serg\_sevbitov@tut.by, tshishkina@tut.by}
\end{center}
\vspace{5mm}

\abstract{The processes of lepton-nucleon scattering, including
ones with both polarized beams, at high energy provide relevant
information about interaction and particles structure, allowing to
analyze nucleon spin structure. As energy and experimental accuracy
rise, necessity to improve Born cross sections and polarized
asymmetries with higher order radiative corrections becomes
substantial. In this report we stress on lowest order bremsstrahlung
corrections treatment using helicity amplitudes method as applied to
actual nowadays charged current lepton-nucleon deep inelastic
processes, that allows to simplify matrix element calculation
procedure. Real photon emission contribution is calculated by means of Lorentz-invariant formalism. Kinematical peculiarities on bremsstrahlung correction are discussed.}

\section{Introduction}

The processes of deep inelastic lepton-nucleon scattering (DIS) are of interest at
present and planned experiments, today with particular emphasis on
both polarized beams interaction investigation, as it provides
essential data on the internal structure of the nucleon spin.
Special interest to charged current interaction is connected with
the absence of large electromagnetic effects contribution to these
processes. Some extensive reviews on nowadays and forthcoming
experimental facilities on such processes can be found for instance
in refs. \cite{hera1,hera2}.
%
Asymmetries withdrew from phenomenological interaction parameters on
certain experiments allow to extract detailed information on
nucleon's spin, concealed in polarized structure functions $g_1$ and
$g_{2(5,6)}$ or individual quark contributions to nucleon's spin. As
expected, obtained information can be used to expand and to refine
nucleon nature knowledge, to compare experimental data with other
related experiments on nucleon structure (e.g. neutral current or
pure electromagnetic DIS ones, which have been studied at a stretch
of many years, see for instance refs. [3--5]) as well as with
Standard Model predictions or perhaps to search deviations from it.

    Processes in question have been investigated before mainly at Born approach, for instance in
our previous papers (see in refs. \cite{sev1,sev2}) we realized Born
level phenomenological analysis in comparison with quark-parton
model approach, Born asymmetry analysis with stress on polarized
structure functions extraction scheme.  In this report we restrict
oneself to detailed treatment of the bremsstrahlung correction
calculation, as correct treatment with observed experimental data at
high energies requires allowance for various radiative effects. Here
to perform calculations we use the formalism of helicity amplitudes
method offered firstly in ref. \cite{ber1,ber2} relevant for single
and multiply bremsstrahlung processes. Using of such analytical
method allows to practically avoid intermediate operations with
traces of Dirac matrices products and undesirable calculations of
cross-elements of $S$-matrix. This method of matrix element
calculation mainly consists in special representation of 4-vectors
of the photon polarization through expressions with bispinors, in
using of special transformation rules likewise Chisholm identities
and in treatment with $\bar{u}_{\mp}(p)u_{\pm}(k)$ constructions as
simple scalar function of $p$ and $k$ to cancellate unnecessary
terms.

\section{Radiative corrections}

 To calculate radiative corrections we employ quark-parton
model, which allows to obtain reasonable quantitative predictions
for nucleon and leptonic bremsstrahlung contributions. Feynman
diagrams of processes in question
\begin{eqnarray*}
l(\bar l) + N \to \nu (\bar \nu ) + X,\\
\nu (\bar \nu ) + N \to l(\bar l) + X,
\end{eqnarray*}
\noindent are presented in FIG.1 (one of the diagram in particular
case vanishes, as neutrino contain no charge).

\begin{figure}[h!]
     \leavevmode
\centering
\includegraphics[width=.43\textwidth]{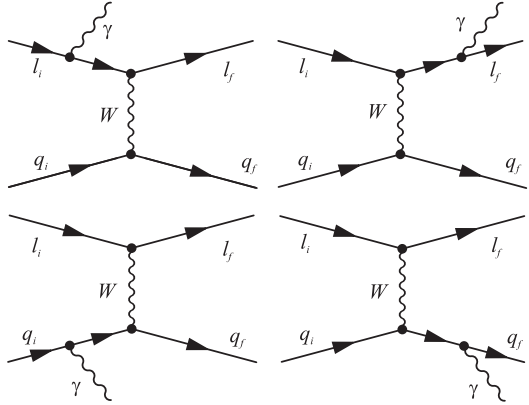}
\caption{Photon bremsstrahlung diagrams for charged current
lepton-nucleon DIS processes.}
\end{figure}

Let's consider firstly the case of $l_i = l,\; l_f = \nu_l ,\;
q_i,\; q_f $. To make use of helicity amplitudes method one should
represent photon polarization vectors as following:
\[ \hat \varepsilon ^ \pm   = \hat \varepsilon _q^ \pm   = N_q
\left[ {\hat q''\hat q'\hat k(1 \mp \gamma _5 ) - \hat k\hat q''\hat
q'(1 \pm \gamma _5 )} \right],\]\[
\quad N_q  = \left( {4\sqrt { - (q''q')(q''k)(q'k)} } \right)^{ -
1}, \]\[
\varepsilon ^ \pm   = 2N_q \left[ {\hat q''\hat q'\hat k(1 \mp
\gamma _5 ) - \hat k\hat q''\hat q'(1 \pm \gamma _5 ) \mp}
\right.\]\[
\left. {\mp \varepsilon _{\nu \alpha \beta \gamma } q''_\alpha
q'_\beta k_\gamma  } \right], \] \noindent where
${q',q'',p'_i,p''_f}$ correspond to incoming $l$ or outgoing $\nu_l$
leptons and quarks $q_i\,q_f$ momenta, $k$ is emitted photon
momentum. This representation conserves common requirements $
\varepsilon ^ \pm  \varepsilon ^ \pm   = 0,\; \varepsilon ^ \pm
{\varepsilon ^ \pm }^ *   = \varepsilon ^ \pm  \varepsilon ^{ \mp  *
} = 1,\; \varepsilon ^ \pm  k = 0$. Given form suitable for
calculation of the leptonic bremsstrahlung term. For hadronic terms
one should involve form, dependent on quark's momenta as following
\[
  \hat \varepsilon ^ \pm   = \hat \varepsilon _p^ \pm   = e^{ \pm i\,\varphi } N_p \left[ {\hat p''_f \hat p'_i \hat k(1 \mp \gamma _5 ) - \hat k\hat p''_f \hat p'_i (1 \pm \gamma _5 )} \right] + \]\[ + \beta _ \pm  \hat k, \]\[
  N_p  = \left( {4\sqrt { - (p''_f p'_i )(p''_f k)(p'_i k)} } \right)^{-1} ,
\]
\noindent where
\[
e^{ \pm i\,\varphi }  = (\varepsilon _q^ \pm  \varepsilon _p^ \mp  )
= \frac{1} {4}Sp\left( {\hat \varepsilon _q^ \pm  \hat \varepsilon
_p^ \mp  } \right) =\]\[ N_p N_q Sp\left[ {\hat p''_f \hat p'_i \hat
k\hat q''\hat q'\hat k(1 \mp \gamma _5 )} \right].
\]
Free parameter $\beta$ can be omitted, as longitudinal component.

  One can get the following expressions
for matrix element using technique thoroughly described in ref.
\cite{ber2}
\[M_{
   -  -  +  -  - }  = - 8\frac{{AN_p e^{ + i\,\varphi } e_f }} {{\sqrt {2k^0 p'_{i0}
p''_{f0} } }}\bar u  (q'')\hat q'u  (p'_i )\bar u   (p''_f )\hat
p'_i \times u  (q')D^W ,
\]\[M_{
   -  -  -  -  - }  =  - 8\frac{{A\left[ {N_q  + N_p e^{ - i\,\varphi } e_i } \right]}}
{{\sqrt {2k^0 p'_{i0} p''_{f0} } }}\bar u  (q'')\hat p''_f u  (p'_i
) \times
\]\[ \times
\bar u  (p''_f )\hat q''u (q')D^W ,\quad \left| A \right|^2  =
\frac{{e^2 G_F^2 M_W^4 }} {2}\frac{1} {{4q'_0 q''_0 }}.
\] \noindent
Here $D^W$ -- $W$-boson propagator, signs $\pm$ refer to the
particle helicity and photon polarization in the following order:
$l,\nu_l,\gamma,q_i,q_f$; $e_i$ and $e_f$ -- initial and final
quark's charges.

     The advantage of using helicity
amplitudes method is that it allows to obtain
 squared matrix element directly without interference terms common in
straightforward calculations. It is readily to show, that squared
matrix element of the lepton-quark process $l q_i \to \nu _l
 q_f $ with real photon emission have the following form
\[
  \left| {M_{--+--} } \right|^2  + \left| {M_{-----} } \right|^2  = \frac{{4 e^2 G_F^2 }}
{{q'_0 q''_0 k_0 p'_{i0} p''_{f0} }}\times\]\[
\times\frac{{M_W^4 }} {{(Q^2  + M_W^2 )^2 }}\left\{
{\frac{1}{2}\left[ {e_i (q'k)(q''k)\left( {\frac{{q'}} {{q'k}} -
\frac{{q''}} {{q''k}}} \right) + }\right.}\right. \]\[\left. + (p'_i
k)(p''_f k)\left( \frac{p'_i }{p''_i k} - \frac{p''_f }{p''_f k}
 \right) \right]^2
\frac{{(p''_f q'')^2 }} {{(p'_i k)(p''_f k)(q'k)(q''k)}}- \]\[
\left.{- \frac{{e_f (q'q'')(p'_i q')^2 }} {{(p'_i k)(p''_f k)}}}
\right\}.
\]

Similar formulas for other cases of electroweak lepton-quark
scattering can be evolved using transformation rules
\[\begin{array}{ccc}
  l\bar q_i  \to \nu _l \bar q_f :\quad p'_i \leftrightarrow p''_f,\quad e_i  \leftrightarrow e_f  \hfill \\
  \bar l{\kern 1pt} q_i  \to \bar \nu _l q_f :\quad q' \leftrightarrow q'', \hfill \\
  \bar l{\kern 1pt} \bar q_i  \to \bar \nu _l \bar q_f :\quad q' \leftrightarrow q'',\quad p'_i \leftrightarrow p''_f,\quad e_i  \leftrightarrow e_f . \hfill \\
\end{array}\]
    We use here the following common notations for kinematical variables:
\[\left\{\begin{array}{ll}
  Q_l  = q' - q'',\quad Q_l ^2  \approx  - 2q'q'', \\
  Q_h  = Q_l  - k = p''_f  - p'_i ,\quad Q_h ^2  \approx  - 2p'_i p''_f ,
  \end{array} \right.\]\[
  \left\{\begin{array}{ll}
  X_i  =  - 2p'_i q'',  \\
  S_i  =  - 2p'_i q',  \end{array} \right.\]\[
  \left\{\begin{array}{ll}
  u =  - 2p'_i k,\quad z_2  =  - 2q''k, \\
  v =  - 2p''_f k = u - Q_l ^2  + Q_h ^2 ,\\
  z_1  =  - 2q'k = z_2  - Q_l ^2  + Q_h ^2 ,\end{array} \right.
\]
 \noindent where $x_{h[l]}= - Q_{h[l]}^2 /2p'Q_h $,
$y_{h[l]} = -2p'Q_{h[l]}/S$
 -- standard hadron and lepton scaling variables, $p'$
and $p''$ -- nucleon and jet 4-momenta.
 To obtain
cross sections or polarized  asymmetry including radiative
corrections one should switch from lepton-quark interaction to
lepton-nucleon one integrating over quark momenta being carried in
the nucleon, and over radiated photon momentum. If we suppose quark
to possess momentum $p'_i=x_{ih} p'$ with the probability of
$f(x_{ih})$, the first integration over $p'_i$ could be performed by
means of the following substitutions:
\[
  f(x_{ih} ) \to \frac{{f(x_h )}}
{{ - 2p'Q_h }} = \frac{{f(x_h )}} {{y_h S}},
 S_i  \to x_{ih} S \to
x_h S = \frac{{Q_h^2 }} {{y_h }},\]\[
  X_i  \to x_{ih} X \to x_h X = \frac{{Q_h^2 }}
{{y_h }}(1 - y_h )S,\]\[\ u \to x_{ih} u \to x_h u,\quad v \to x_h u -
Q_l^2  + Q_h^2 ,
\]
keeping  $Q_h ^2  = (Q_l  - k)^2$, $Q_l ^2$, $z_1$ and $z_2$
unaltered. Here $S =  - (p' + q')^2$.

    To integrate over photon momentum one can use
covariant method of integration described, for instance, in ref.
\cite{sh1,sh2}, permitting to integrate directly over
Lorentz-invariant kinematical variables.
 Covariant calculation has advantage of missing of the sophisticated Monte-Carlo techniques but presence of  the analytical integration as well as it can be carried out for various   kinematical
experimental configurations.

  Firstly, lets imply the following suitable phase space
transformation, allowing to derive from its general form the expression
containing introduced before invariant variables:
\[
 d\Gamma  = dM_h^2 \frac{{d^3 p''}} {{2p''_0 }}\frac{{d^3 q''}}
{{2q''_0 }}\frac{{d^3 k}} {{2k_0 }}\delta ^{(4)} (Q_l  - k - Q_h )
=
\]
\[
=dM_h^2 dQ_h^2 \frac{{d^3 q''}} {{2q''_0 }}\frac{{d^3 k}} {{2k_0
}}\delta \left[ {(Q_l  - k)^2  + M_h^2 } \right] \times \]
\[\times\delta
\left[ {Q_h^2  - (p'' - p')^2 } \right] = \frac{{\pi S}} {2}dy_l
dQ_l^2 dy_h dQ_h^2 \frac{{dz}} {{2\sqrt {R_z } }}.
\]
Here $R_z$ is the Gram determinant \cite{byck} of 4-vectors $q'$,
$p'$, $q''$, $p''$
\[R_z =  - \Delta_4(q', p', q'', p'')=\]\[=-\left|\begin{array}{cccc}q'^2&q'p'&q'q''&q'p''\\ p'q'&p'^2&p'q''&p'p''\\q''q'&q''p'&q''^2&q''p''\\p''q'&p''p'&p''q''&p''^2\end{array}\right|,\]
\noindent which can be expressed as quadratic polynomial of $z_1$ or
$z_2$ variables defined before
\[
 R_z = -A z^2 + 2 B z - C,
\]
\noindent where the coefficients in the ultrarelativistic limit are
\[
A_{1,2}=y_l^2 S^2 + 4 M^2 Q_l^2,\]
\[
B_1=-2 M^2 Q_l^2 (Q_l^2 - Q_h^2) + (y_l Q_h^2 - y_h Q_l^2) S^2 +
\]
\[+(1 - y_l) S^2 Q_l^2 (y_l - y_h) - m^2 (2 M^2
 Q_h^2 + 2 M^2 Q_l^2 - S^2 y_h y_l),\]
\[
B_2=2 M^2 Q_l^2 (Q_l^2 - Q_h^2) + (1 - y_l)(y_l Q_h^2 - y_h Q_l^2)
S^2 +\]
\[+ S^2 Q_l^2 (y_l - y_h) -  m^2(2 M^2
 Q_h^2 +2 M^2 Q_l^2 - S^2 y_h y_l),\]
\[
C_1=S^2[Q_h^2 + (-1 + y_l - y_h) Q_l^2]^2 +\]
\[+ 4  m^2 Q_l^2 (y_l - y_h)^3 (1 - y_l),\]
\[ C_2=S^2[(1 - y_l) Q_h^2 - (1 - y_h) Q_l^2]^2 +\]
\[+ 4  m^2 Q_l^2 (y_l - y_h)^3 (1 - y_l)^{-1}.
\]
 In presented above expression we simplified common phase space
by means of auxiliary invariant variables $z_1$ or $z_2$. Next one
can employ the following integration scheme:
\[
  d\sigma \sim \int\limits_{y_{h\min } }^{y_{h\max } } {dy_h \int\limits_{Q_{h\min }^2 }^{Q_{h\max }^2 } {dQ_h^2 \int\limits_{z_{\min } }^{z_{\max } } {\frac{{dz\, dy_l\, dQ_l^2}}{{y_h S \sqrt {R_z } }} A}}},
\]
or
\[ d\sigma \sim \int\limits_{y_{l\min } }^{y_{l\max } } {dy_l
\int\limits_{Q_{l\min }^2 }^{Q_{l\max }^2 } {dQ_l^2
\int\limits_{z_{\min } }^{z_{\max } } {\frac{{dz\, dy_h\, dQ_h^2}}
{{y_h S \sqrt {R_z }}} A }}},
\]
\[A= \left| {M} \right|^2  f_i (x_h, Q_h^2),\]
\noindent and so on, dependently on desired final variables. Here matrix element
$\left| {M} \right|^2$ expressed in terms of $Q_{l[h]}^2,y_{l[h]}$
and $z_{1[2]}$ have the following form:
\[
   \left| {M(S_i ,X_i ,Q_h ,Q_l ,z_{[1,2]} )} \right|^2  \sim \]\[\frac{{e_f^2 S_i^2 Q_l^2 }}
{{2uv}} + \mathop {\left( {z_2  - Q_l^2  - X_l } \right)}\nolimits^2  \times  \]\[
   \times \frac{{\left[ {Q_h^2 uv +  \left( {e_i^2 Q_l^2 z_1 z_2  - e_i Q_l^2 u (z_1  + z_2 ) + e_i Q_l^2(u - v)(S_i z_2  - X_i z_1 ) }\right)} \right]}}
{{2uv z_1 z_2 }}
\]
for $l_i  = l,\; l_f  = \nu _l ,\; q_i ,\; q_f $ and
\[
   \left| {M(S_i ,X_i ,Q_h ,Q_l ,z_{[1,2]} )} \right|^2  \sim \]\[ \frac{{Q_h^2 X_l^2 }}
{{2z_1 z_2 }} + \frac{{e_f^2 Q_l^2 X_l^2 }} {{2uv}} + \frac{{\mathop
{e_i^2 Q_l^2 \left( {z_1  + Q_l^2  - S_i } \right)}\nolimits^2 }}
{{2uv}} +  \]\[
   + \frac{{e_f X_l^2 \left[ {uQ_l^2 (z_1  + z_2 ) - (u - v)(S_i z_2  - X_i z_1 )} \right]}}
{{2uvz_1 z_2 }}\]
for $l_i  = l,\; l_f  = \nu _l ,\;\bar q_i ,\;\bar q_f $.

 Calculation of the integral over $z$ can be carried out using simple
table integrals of the form
\[I=\int\limits_{z_{min}}^{z_{\max}}{\frac{z^n}{A(z-z_{\min})(z_{\max}-z)}{dz}}, \quad n=-2\ldots 2.\]
To evaluate remaining integrals one should firstly use some
parameterizations on quark distribution functions $f_i (x_h, Q_h^2)$,
for instance QCD-based fits from ref. \cite{alekh2}, and then
choose final variables in which result cross-section or asymmetry
will be expressed.

 In order to calculate infrared contribution, one can apply the
limits  $ \mathop {\lim }\limits_{k \to 0} (u - v) = 0 $ and $
\mathop {\lim }\limits_{k \to 0} (z_2  - z_1 ) = 0 $ to presented
above relations. When calculating this part requires using of some
regularization method as being infrared divergent part, for instance
by introducing virtual photon mass or applying dimensional
regularization method. Here we didn't stress on soft photon emission
contribution calculation, as virtual loop contribution and soft
emission one, infrared divergent separately, compensate partly each
other, remaining uncompensated part have minor influence on the
asymmetries for processes in question as reduced factor.

 Kinematical peculiarities and variables limits for these integrals are thoroughly described e.g. in ref. \cite{byck}, here we give only kinematical relations, necessary for determination of the integration bounds. 
Imposed constrains on the physical region of invariants have the
following form in terms of kinematical $\lambda$-functions (see
\cite{byck})
\[\lambda(\lambda_S,\lambda_l,\lambda_q)\leq 0, \;
\lambda(\lambda_q,\lambda_h,\lambda_k)\leq 0,\]
\[
\lambda(\lambda_S,\lambda_\tau,\lambda_h)\leq 0,\]
\[
\lambda_k=(y_l-y_h)^2 S^2,\; \lambda_S=S^2-4 m^2 M^2,\]
\[
\lambda_l=(1-y_l)^2 S^2, \; \lambda_q=y_l^2 S^2 +4 M^2 Q_l^2,\]
\[
\lambda_h =y_h^2 S^2 +4 M^2 Q_h^2, \; \lambda_\tau=(1-y_h)^2 S^2 -4
M^2 z_2, \]
\[
\lambda(x,y,z)=x^2+y^2+z^2-2xy -2yz - 2xz.
\]
The boundary equations for this conditions can be expressed in the
form of three equations \noindent
\[
M^2(Q_l^2-m^2)^2+Q_l^2 y_l S^2 - m^2 S^2 y_l(1-y_l)-Q_l^2 \lambda_S = 0,\]
\[
y_l^2 S^2 Q_h^2 + y_h^2 S^2 Q_l^2 -M^2 (Q_l^2 -Q_h^2)^2-y_l y_h (Q_l^2 +Q_h^2)S^2 = 0 ,\]
\[
(2M^2 z_2^2-2 m^2 M^2+ y_h S^2+ 2 M^2 Q_h^2)^2 -\lambda_S \lambda_h
=0,
\]
\noindent consequently.

 One can obtain certain integration bounds for chosen final
variables by combining these constraints with
$z_{1}^{{\min},{\max}}$ and $z_{2}^{{\min},{\max}}$ emerging from
the condition $R_{z_{1,2}}\geq 0$.

In ref. \cite{sev1,sev2} we numerically calculated hard real photon
emission contribution with low photon energy cut parameter
$\varepsilon_{\mathrm{cut}}$, where unpolarized quark distribution
functions from ref. \cite{alekh2} with polarized one from ref.
\cite{lss2005} were taken. In FIG.2 we cite comparison of the
corrected polarized asymmetry \noindent
\[ A_{\parallel}
= {\left( \frac{d^2 \sigma ^{{\begin{subarray}{l}
   \uparrow  \Uparrow  \\
\end{subarray}}{}}}{dxdy}-\frac{d^2 \sigma
^{{\begin{subarray}{l}
   \uparrow  \Downarrow  \\
\end{subarray}}{}}}{dxdy}\right)}
{/\left( \frac{d^2 \sigma ^{{\begin{subarray}{l}
   \uparrow  \Uparrow  \\
\end{subarray}}{}}}{dxdy}+\frac{d^2 \sigma
^{{\begin{subarray}{l}
   \uparrow  \Downarrow  \\
\end{subarray}}{}}}{dxdy}\right)},
\]
\noindent $d^2\sigma/dxdy$ -- differential cross section,
$\uparrow(\downarrow)$ corresponds to lepton helicity value $-1(1)$,
$\Uparrow(\Downarrow)$ for nucleon spin, parallel (antiparallel) to
lepton momentum, with the Born one in lepton scaling variables $x_l$
and $y_l$. As for relative contribution, it increases with $x_l$ growth,
reaching values for $x_l=0.9$  approximately larger by 20 \% than
for $x_l=0.01$; it roughly inversely exponentially dependent on
$y_l$, not exceeding $\approx 30\%$ for $y_l>0.3$ and $x_l<0.9$,
$\approx 20\%$ for $y_l>0.3$ and $x_l<0.25$ and $\approx 10\%$ for
$y_l>0.3$ and $x_l<0.1$.
 Asymmetry absolute values depend on $y_l$ noticeable unlike Born ones.

    As for multiply bremsstrahlung, such contributions can be obtained
on the basis of single one by applying renormalization group
equations, and will be treated hereafter.

In conclusion, real photon bremsstrahlung contribution calculated
here affects significantly cross-sections and polarized asymmetries,
so it's necessary to take into account such contribution, in that
way it can be used for certain future experimental needs with the
aim of precision DIS data extraction to improve accuracy of quark
distribution functions as well as to detail nucleon's spin
structure.
\begin{figure}
\centering
\includegraphics[width=.7\textwidth]{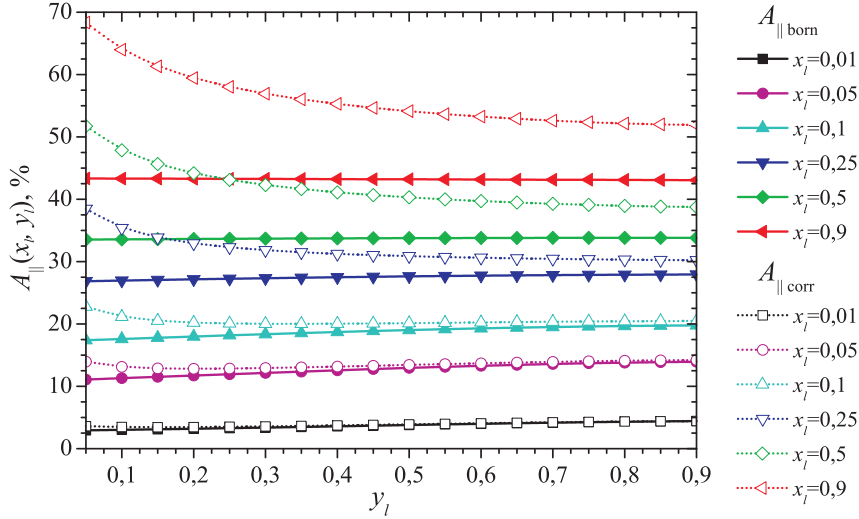}
\caption{Born (solid line) and next to Born order (dotted line)
longitudinal polarized asymmetry $A_\parallel (x_l, y_l)$ for
$lN\rightarrow\nu_l X$ at $E=100$ GeV and
$\varepsilon_{\text{cut}}=1\text{ MeV}$.}
\end{figure}
%
%

%



\end{document}